# Physics Education using a Smartphone Accelerometer

Randall D. Peters

Physics Department
Mercer University, Macon, GA

**Abstract**

Described is an experiment in which a smartphone was caused to move at steady state in a vertical plane, on a path that was nearly circular. During a time interval of data acquisition that encompassed multiple orbits, the acceleration of the phone was measured by means of its internal accelerometer. A subsequent analysis of the data that was collected shows reasonable agreement between experiment and a simple theory of the motion.

**Background**

In the last decade great attention has been given to methods for improving physics education. Some have devoted themselves to theory including thought experiments. Their efforts have been most impressive when the problems studied were amenable to computer modelling. Others have focused on hardware innovations, recognizing that conceptual understanding can be greatly improved through effective hands-on experience in the laboratory. When possible, the ideal teaching tool provides student involvement with a package comprising both theory and experiment. The hardware presently described is a good candidate with which to develop experiments of this type. It is a recently marketed smartphone, the Droid, which should prove useful in the study of certain topics in Newtonian mechanics.

**Hardware**

An accelerometer is part of the internal hardware of the Droid-x smartphone that was used in this study [1]. To control this accelerometer, a `seismo app' [2] was downloaded and installed in the operating system of the phone.

The same leather belt used to presently `twirl' the phone was previously used in a pendulum study [3]. For the present experiment, the phone was placed in a loop at one end of the belt, and with a hand at the other end it was swung around in a vertical orbit, the trajectory of which was approximately circular. The orbit could not be a perfect circle because the hand dynamic required to sustain the motion approximates a small circle of its own. Nevertheless, as will be seen from the results that follow, the phone's motion is nearly that of a pendulum undergoing steady rotation with ever increasing angular displacement.

**Experiment**

For data acquisition, the accelerometer program was initialized, and then motion of the phone was established. Ignoring the transient immediately after phone initialization, the recorded acceleration along the x-axis is shown in Fig. 1. Determined by the phone's orientation in the belt loop, the direction of the x-axis in this case was one pointing vertically downward when the phone was at its lowest point in the orbit. As such, the recorded data corresponds to radial acceleration.

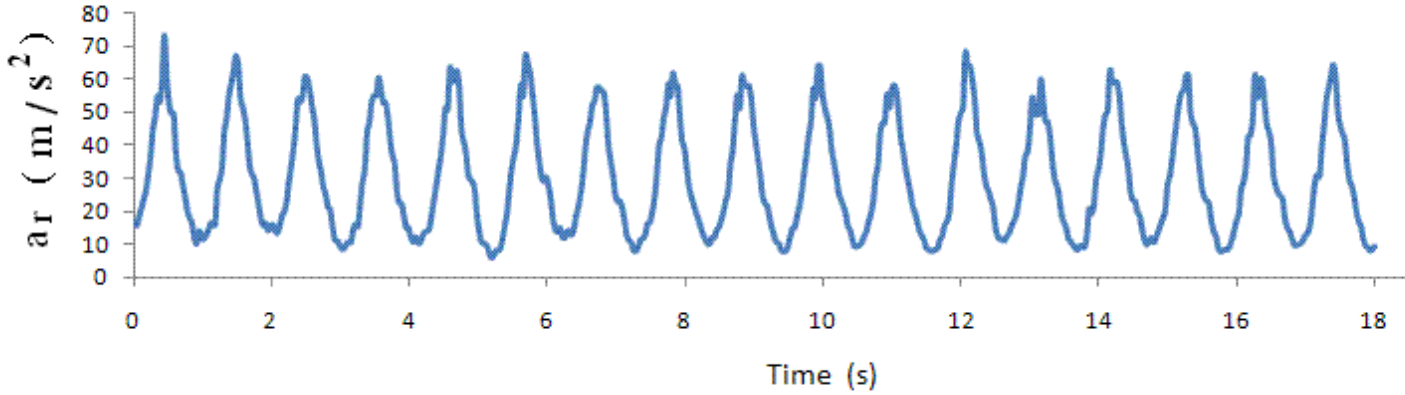

**Figure 1.** Time dependent radial acceleration of the smartphone during seventeen near-circular orbits confined to a vertical plane.

All graphs of this article were generated using Microsoft Excel on a personal computer. The record saved to Droid-x memory was transferred to the computer using gmail.

**Theory**

Theoretical treatment is straightforward using the equation of motion for a simple pendulum; i.e.,

$$\tau = I\alpha = ml^2\ddot{\theta} = -mgl\sin\theta \qquad (1)$$

where $l$ is the length of the pendulum.

To compare theory and experiment, a numerical approximation to the integral of Eq.(1) was performed with Excel, using the last point approximation. In other words, the following equations were updated in the order shown in Eq.(2), using $\delta t = 0.033$ s. This time step corresponds to the reciprocal of the data sampling rate, specified for the seismo-app to be 30 samples per second.

$$\begin{aligned}\omega_{n+1} &= \omega_n - \delta t\,(g/l)\sin\theta \\ \theta_{n+1} &= \theta_n + \omega\,\delta t\end{aligned} \qquad (2)$$

Estimates were obtained by trial and error for the following three adjustable parameters of the motion: (i) the initial angular velocity $\omega_0$, (ii) the initial displacement $\theta_0$, and (iii) the effective radius $l$ of the orbit. The natural order with which to search for the best fit to the data followed that of the parameter listing. The amount of asymmetry in the theoretical trace is determined by $\omega_0$, as illustrated in Fig. 2.

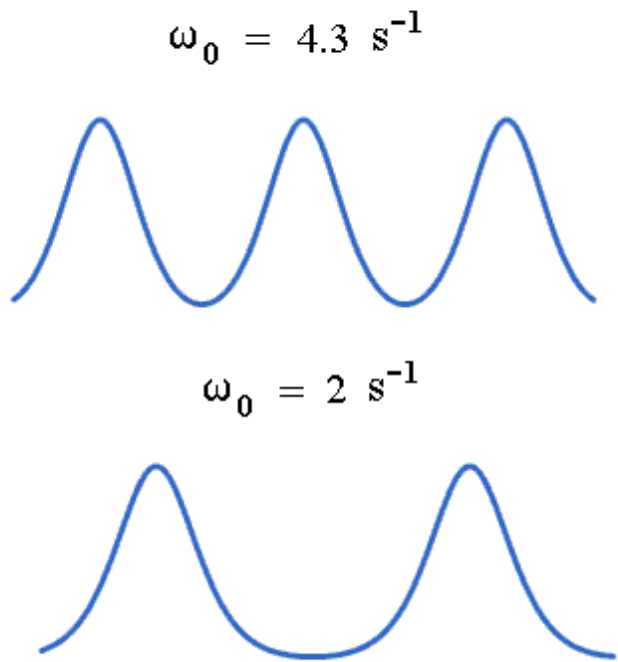

**Figure 2.** Illustration of the dependence of trace-shape asymmetry on $\omega_0$ (which also establishes the fixed, average angle rate).

After selecting $\omega_0 = 4.3$ rad/s as a reasonable approximation to the shape of the experimental trace, the other two parameters were adjusted, also by straightforward (rapid) trial and error using Excel, and finally set at $\theta_0 = 0.4$ rad and $l = 0.76$ m. With these values, the theoretical curve shown as red in Fig. 3 was generated.

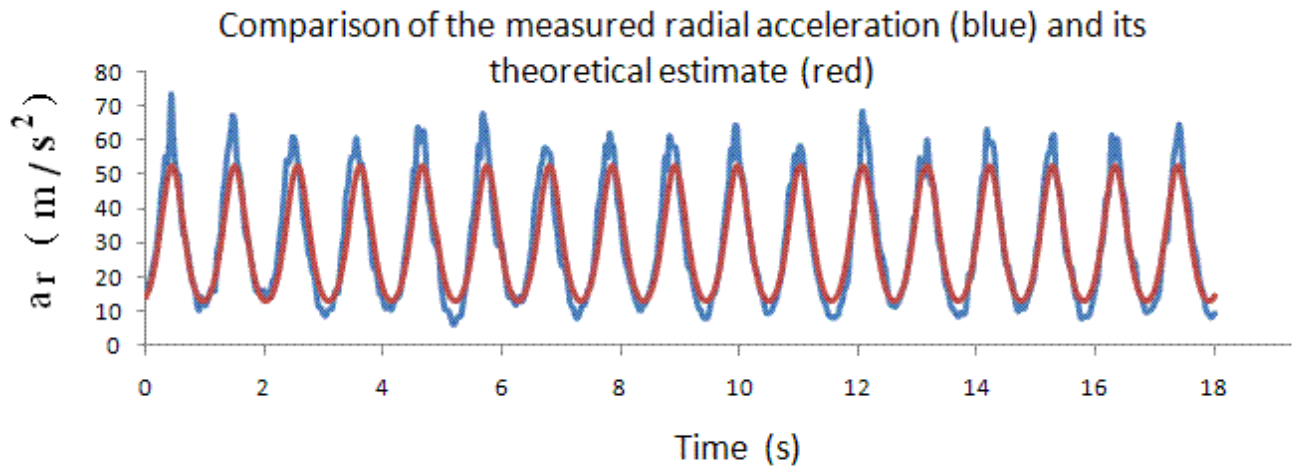

**Figure 3.** Comparison of theory and experiment.

The comparison is reasonable, in light of the `noises' evident in the experimental data. These derive from complex motions associated with both (i) the author's inability through hand motion to establish and maintain a pure (repeatable) orbit, and (ii) twisting and other mechanical modes of the belt.

It is interesting to look at the spectra of the curves shown in Fig. 3. They are readily generated by the FFT algorithm (named Fourier analysis in Excel), and are shown in Fig. 4. Asymmetry of the acceleration is responsible for the significant harmonic content that is visible in the plots. What was plotted for these graphs is the `single-sided' transform, in which ordinate values are equal to twice the square of the modulus of the FFT components. The Excel output is not normalized and useful for power spectral density graphing until after the raw values are divided by the square of the number of points N in the FFT set. The articles cited in references 4 and 5 explain why $N = 2^n$, where n is an integer. Here N = 512, so that the first 256 points were used to generate the plots. The unused second half of the set corresponds to negative frequencies and is not used in the single-sided transform. For Parseval's theorem to be satisfied, the half-set values must be multiplied by a factor of two.

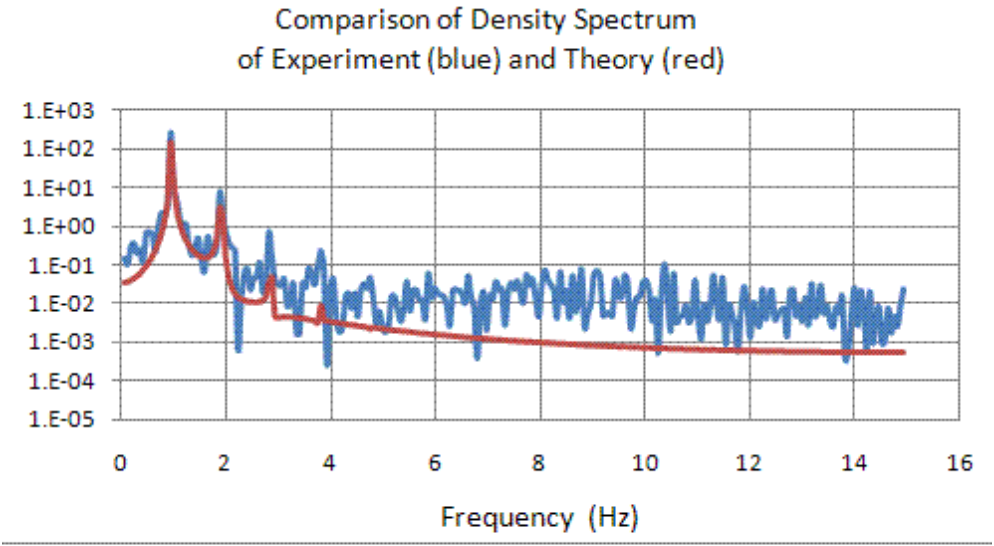

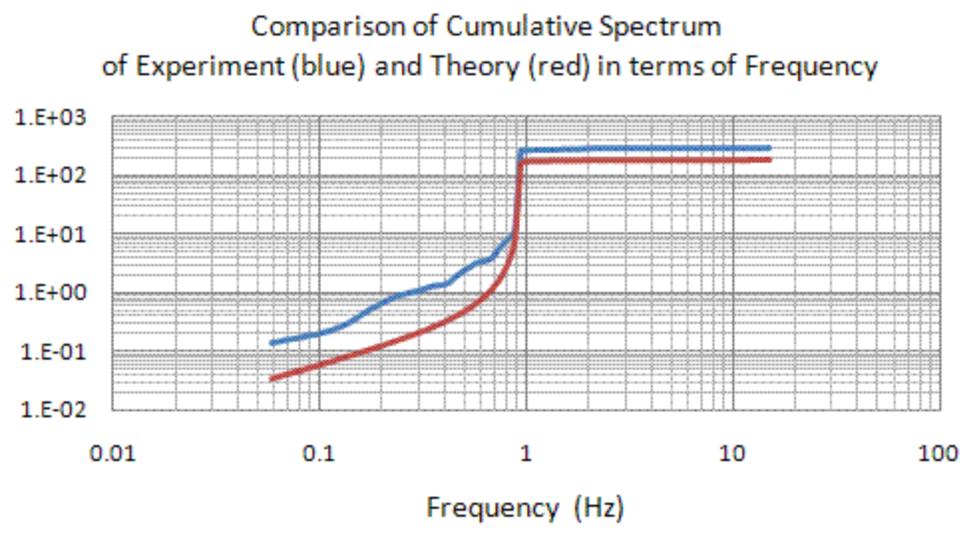

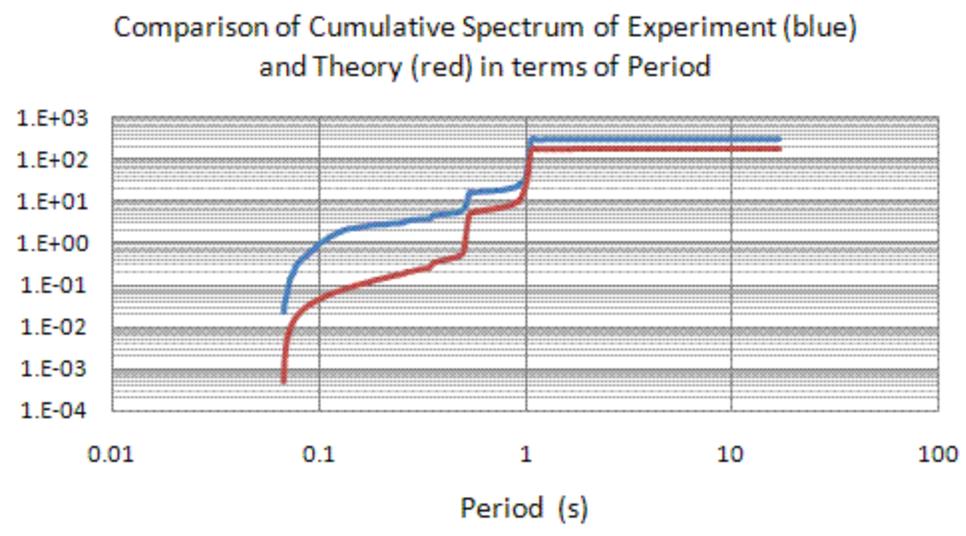

**Figure 4.** Spectra of theory and experiment for the curves shown in Fig. 3.

The cumulative spectrum is obtained from the density spectrum by integrating over the latter [6]. Because it is less `cluttered', it is amenable to the overlay of multiple plots in a single graph.

**Conclusion**

The present experiment, along with the previous pendulum experiment [3], shows that the Droid accelerometer could open a host of student-friendly laboratory or playground activities. For example, an obvious formal laboratory exercise would be one in which the smartphone rides on an air-track cart executing simple harmonic motion. Fun-time playground activities could involve students engaging themselves with various rides of type found in parks, with the smartphone secured in their pocket. No doubt the reader can envision a variety of other possibilites, such as carnival rides.

[1] Manufactured by Motorola, distributed by Verizon.
[2] Program developed by Ari Wilson, available without charge in the `tools' category of the Verizon `Market'.
[3] Peters, R. D. (2010), ``Smart-phone sensor of pendulum motion'', online at http://arxiv.org/abs/1012.1800
[4] Peters, R. D. (1992), ``Fourier transform construction by vector graphics'', Amer. J. Phys. 60, 439.
[5] Peters, R. D. (2003), ``Graphical explanation for the speed of the Fast Fourier Transform'', online at http://arxiv.org/html/math/0302212v1
[6] Lee, S. C. & Peters, R. D. (2009), ``A new look at an old tool-the cumulative spectral density of Fast Fourier Transform Analysis'', online at http://arxiv.org/abs/0901.3708 also Peters, R. D. (2007), ``A new tool for seismology-the Cumulative Spectral Power, online at http://arxiv.org/abs/0705.1100